# QUBO Refinement: Achieving Superior Precision through Iterative Quantum Formulation with Limited Qubits

Hyunju Lee
Quantum Research Center
QTomo Inc.
Cheongju, Republic of Korea
hyunjulee0824@gmail.com

Kyungtaek Jun
Quantum Research Center
QTomo Inc.
Cheongju, Republic of Korea
Chungbuk Quantum Research Center
Chungbuk National University
Cheongju, Republic of Korea
ktfriends@gmail.com

*Abstract*— **In the era of quantum computing, the emergence of quantum computers and subsequent advancements have led to the development of various quantum algorithms capable of solving linear equations and eigenvalues, surpassing the pace of classical computers. Notably, the hybrid solver provided by the D-wave system can leverage up to two million variables. By exploiting this technology, quantum optimization models based on quadratic unconstrained binary optimization (QUBO) have been proposed for applications, such as linear systems, eigenvalue problems, RSA cryptosystems, and CT image reconstruction. The formulation of QUBO typically involves straightforward arithmetic operations, presenting significant potential for future advancements as quantum computers continue to evolve. A prevalent approach in these developments is the binarization of variables and their mapping to multiple qubits. These methods increase the required number of qubits as the range and precision of each variable increase. Determining the optimal value of a QUBO model becomes more challenging as the number of qubits increases. Furthermore, the accuracies of the existing Qiskit simulator, D-Wave system simulator, and hybrid solver are limited to two decimal places. Problems arise because the qubits yielding the optimal value for the QUBO model may not necessarily correspond to the solution of a given problem. To address these issues, we propose a new iterative algorithm. The novel algorithm sequentially progresses from the highest to the lowest exponent in binarizing each number, whereby each number is calculated using two variables, and the accuracy can be computed up to a maximum of 16 decimal places.**

## I. INTRODUCTION

With the advent of quantum computing, the landscape of computational paradigms has undergone a profound transformation [1]. Based on the manipulation of various quantum states, such as superposition, entanglement, and interference, quantum computing has the potential to surpass the capabilities of classical computers by leveraging quantum–mechanical phenomena. [2,3]. During the late 1990s, notable quantum algorithms in factorization, quantum searches, and prediction of linear equation solutions were introduced, leading to a diverse array of promising applications and unparalleled computational speed [4,5,6]. Their influence resonates across various business and research domains, creating new possibilities and advancements. Quantum machine learning (QML) is a promising frontier in quantum computing with an ability to generalize intricate patterns, which eludes its classical counterparts. Recently, several efforts have been made to integrate quantum algorithms into the domain of machine learning [7-11].

The quantum annealing processor, which is a notable quantum computing instantiation, inherently provides solutions characterized by low-energy states. Leveraging quantum annealing for problem resolution necessitates the formulation of quadratic unconstrained binary optimization (QUBO) representations that transform problems into quests for minimum values. This nondeterministic polynomial time (NP) mathematical framework has prompted a surge in research directed at resolving various NP-hard problems using QUBO formulations [12-15] in linear regression, support vector machines, and clustering, rendering this framework suitable for QML training on adiabatic quantum computers [16-21].

Even in numerical computation, which is an indispensable facet of scientific and engineering domains, the escalating volume of data underscores the imperative for expeditious solutions to large-scale problems. In 2009, Harrow et al. introduced a seminal quantum algorithm for linear system solutions on gate-model quantum computers [22]. This algorithm offers exponential acceleration for sparse and well-conditioned linear systems, prompting subsequent endeavors to enhance linear system solving based on its principles [23-27]. A recent contribution by Jun and Lee presents numerical QUBO formulation methods for solving linear systems and eigenvalue problems with general n × n matrices on a quantum annealer [28,29]. Notably, these methods harness the parallel computing capabilities intrinsic to the QUBO modeling process, to substantially accelerate computations. However, the QUBO formulation for solving the prime factorization problem on a quantum annealer [30], has the potential to threaten the security of the Rivest–Shamir–Adleman (RSA) cryptosystem, which relies on the difficulty of factoring large numbers. Jun recently developed quantum algorithms for computed tomography (CT) image reconstruction [31]. The newly proposed quantum optimization algorithm for CT image reconstruction is expected to generate superior images compared to those created using conventional iterative methods as it calculates the global energy minimum/maximum. This algorithm is anticipated to contribute significantly to the advancements in the field of medical imaging and holds promise for substantial progress.

In addition to advancements in quantum annealing, quantum optimization algorithms are undergoing significant progress. The quantum processing unit (QPU) solver provided by D-Wave offers approximately 180 logical qubits, whereas the hybrid solver supports up to 2 million variables. The hybrid solver can compute a QUBO matrix of up to 2,000,000 × 2,000,000, which showcases its potential for solving complex problems across various domains.

Furthermore, with the advancement of parallel quantum annealing methods, more efficient quantum optimization computations are anticipated [32,33,34]. However, as the dimensions of the QUBO matrix increase, the number of coefficients also increases, potentially leading to larger physical errors in quantum computations. The bias and coupler of a quantum annealer typically have a physically implementable range with an accuracy of approximately $O(n^{-2})$ [35].

In this study, we demonstrate a method that uses a hybrid solver to compute the accuracy of the energy model up to a maximum of $O(n^{-25})$. The novel algorithm reduces both the global minimum energy required for solving a given problem and the number of coefficients involved in the computation. This approach is applicable to quantum optimization algorithms that approximate solutions to a given problem by monitoring the global minimum/maximum energy. Moreover, representing variables as a superposition of binary states allows adjusting the number of logical qubits used in computing the problem. To validate the new algorithm, we utilize a hybrid solver to compute only a single solution for a linear system comprising irrational numbers. Using the new algorithm, solutions of up to 16 decimal places were accurately determined for a linear system. The algorithm demonstrates a method for overcoming the physical limitations of couplers and bias within the algorithm, helping discover more effective solutions in quantum optimization models.

## II. METHOD

### A. Background

The algorithm introduced in this study can be used for any form representing a variable as a linear combination of qubits. To validate this algorithm, we applied it to a linear system. First, we introduce the quadratic unconstrained binary optimization (QUBO) model [36, 37] for linear systems. QUBO has emerged as a pivotal challenge in combinatorial optimization, extending its reach across diverse applications in industry [38] with various computer science problem domains systematically embedding into QUBO [39]. The resolution of problems employing quantum annealers entails transforming these challenges into Ising functions expressed in terms of logical variables or an equivalent QUBO format. Subsequently, seamless integration of the logical problem into the physical architecture of the quantum annealer is achieved through meticulous mapping of logical variables onto qubits.

Quantum annealing algorithms, grounded in the exploitation of quantum effects navigate local minima through tunneling phenomena, to identify the global minimum within a given cost function [40]. Within the context of this particular challenge, the cost function denoted as f is defined within an n-dimensional binary vector space, $\mathbb{B}^n$, on real numbers, $\mathbb{R}$.

$$f(\vec{q}) = \vec{q}^T Q \vec{q} \qquad (1)$$

where $Q$ is an upper triangular matrix; $\vec{q} = (q_1, \cdots, q_N)^T$; and $q_i$ is a binary element of $\vec{q}$. Within the scope of this manuscript, matrix $Q$ denotes the QUBO matrix. The objective is to identify the vector $\vec{q^*}$ that minimizes the cost function $f$ within the set of vectors $\vec{q}$. Given the condition $q_i^2 = q_i$, the cost function can be redefined as follows:

$$f(\vec{q}) = \sum_{i=1}^{N} Q_{i,i} q_i + \sum_{i<j}^{N} Q_{i,j} q_i q_j \qquad (2)$$

Within matrix $Q$, the diagonal elements $Q_{i,i}$ symbolize the linear terms, whereas off-diagonal elements $Q_{i,j}$ represent the quadratic terms. The variables of the Ising model, denoted as $\sigma$, and those of the QUBO model, denoted as $q$, exhibit a linear relationship.

$$\sigma \to 2q - 1 \text{ or } q \to \frac{1}{2}(\sigma + 1). \qquad (3)$$

Therefore, the appropriate format is selected depending on the unknown binary.

For a matrix $A \in R^{n \times n}$, given a column vector of variables $\vec{x} \in R^n$ and a column vector $\vec{b} \in R^n$, the objective is to determine $\vec{x}$ such that $A\vec{x} = \vec{b}$. The objective of the linear least-squares problem is to identify the vector $\vec{x}$ that minimizes the norm $\|A\vec{x} - \vec{b}\|$, which can be expressed as follows:

$$\arg\min_{\vec{x}} \|A\vec{x} - \vec{b}\| \qquad (4)$$

To solve (4) let us express $\|A\vec{x} - \vec{b}\|$:

$$A\vec{x} - \vec{b} = \begin{pmatrix} a_{1,1} & a_{1,2} & \cdots & a_{1,n} \\ a_{2,1} & a_{2,2} & \cdots & a_{2,n} \\ \vdots & \vdots & \ddots & \vdots \\ a_{n,1} & a_{n,2} & \cdots & a_{n,n} \end{pmatrix} \begin{pmatrix} x_1 \\ x_2 \\ \vdots \\ x_n \end{pmatrix} - \begin{pmatrix} b_1 \\ b_2 \\ \vdots \\ b_n \end{pmatrix} \qquad (5)$$

Squaring the 2-norm of the resulting vector from (5), we obtain the following:

$$\| A\vec{x} - \vec{b} \|_2^2 = \vec{x}^T A^T A \vec{x} - 2\vec{b}^T A \vec{x} + \vec{b}^T \vec{b} \qquad (6)$$

$$= \sum_{k=1}^{n} \left\{ \left(\sum_{i=1}^{n} a_{k,i} x_i\right)^2 - 2b_k \sum_{i=1}^{n} a_{k,i} x_i + b_k^2 \right\} \qquad (7)$$

$$= \sum_{k=1}^{n} \left\{ \sum_{i=1}^{n} (a_{k,i} x_i)^2 + 2 \sum_{i<j} a_{k,i} a_{k,j} x_i x_j - 2b_k \sum_{i=1}^{n} a_{k,i} x_i + b_k^2 \right\} \qquad (8)$$

In the context of solving binary linear least squares, each variable $x_i$ is encoded through a combination of qubits $q_{i,j} \in \{0,1\}$. O'Malley and Vesselinov [39] elaborated on the radix-2 representation of the positive real value $x_i$, denoted as

$$x_i \approx \sum_{l=-m}^{m} 2^l q_{i,l} \qquad (9)$$

where a positive integer $l$ denotes the number of digits in the integer part of $x_i$, and a negative $l$ represents the number of digits in the fractional part. The real value $x_i$ is represented as follows.

$$x_i \approx \sum_{l=-m}^{m} 2^l q_{i,l}^+ - \sum_{l=-m}^{m} 2^l q_{i,l}^- \tag{10}$$

or

$$x_i \approx \sum_{l=-m}^{m} 2^l q_{i,l}^+ - 2^{m+1} q_i^- \tag{11}$$

Note that this representation may yield the same value with different binary combinations.

To formulate the QUBO model, (10) is substituted into (8). Then, the terms for the variable $x_i$ in (8), can be expanded using (12), (13) and (14), and expressed in terms of qubits.

$$\sum_{k=1}^{n}\sum_{i=1}^{n}(a_{k,i}x_i)^2 \approx \sum_{k=1}^{n}\sum_{i=1}^{n}\sum_{l=-m}^{m} a_{k,i}^2 \, 2^{2l}(q_{i,l}^+ + q_{i,l}^-) \tag{12}$$
$$+ \sum_{k=1}^{n}\sum_{i=1}^{n}\sum_{l_1<l_2} a_{k,i}^2 \, 2^{l_1+l_2+1}(q_{i,l_1}^+ q_{i,l_2}^+ + q_{i,l_1}^- q_{i,l_2}^-)$$

$$\sum_{k=1}^{n}\sum_{i<j} 2\,a_{k,i}a_{k,j}x_i x_j \tag{13}$$
$$\approx \sum_{k=1}^{n}\sum_{i<j}\sum_{l_1=-m}^{m}\sum_{l_2=-m}^{m} 2^{l_1+l_2+1}\, a_{k,i}a_{k,j}$$
$$(q_{i,l_1}^+ q_{j,l_2}^+ + q_{i,l_1}^- q_{j,l_2}^- - q_{i,l_1}^+ q_{j,l_2}^- - q_{i,l_1}^- q_{j,l_2}^+)$$

$$\sum_{k=1}^{n}\sum_{i=1}^{n}(-2a_{k,i}b_k x_i) \tag{14}$$
$$\approx \sum_{k=1}^{n}\sum_{i=1}^{n}\sum_{l=-m}^{m} 2^{l+1}\, a_{k,i}b_k(q_{i,l}^- - q_{i,l}^+)$$

The QUBO model for a linear system is the sum of the three aforementioned terms.

*B. Iterative QUBO formulations*

The new QUBO model is formulated for the case that each $x_i$ is $2^l(q_{i,l}^+ - q_{i,l}^-)$ in the QUBO model for a linear system with a general solution. The new QUBO model is calculated using (12), (13), and (14), as follows:

$$\sum_{k=1}^{n}\sum_{i=1}^{n}(a_{k,i}x_i)^2 \approx 2^{2l}\left\{\sum_{k=1}^{n}\sum_{i=1}^{n} a_{k,i}^2\left(q_{i,l}^+ + q_{i,l}^-\right)\right. \tag{15}$$
$$\left. - \sum_{k=1}^{n}\sum_{i=1}^{n} 2a_{k,i}^2\, q_{i,l}^+ q_{i,l}^-\right\}$$

$$\sum_{k=1}^{n}\sum_{i<j} 2\,a_{k,i}a_{k,j}x_i x_j \tag{16}$$
$$\approx \sum_{k=1}^{n}\sum_{i<j} 2^{2l+1}\, a_{k,i}a_{k,j}(q_{i,l}^+ q_{j,l}^+ + q_{i,l}^- q_{j,l}^-$$
$$- q_{i,l}^+ q_{j,l}^- - q_{i,l}^- q_{j,l}^+)$$

$$\sum_{k=1}^{n}\sum_{i=1}^{n}(-2a_{k,i}b_k x_i) \approx \sum_{k=1}^{n}\sum_{i=1}^{n} 2^{l+1}\, a_{k,i}b_k(q_{i,l}^- - q_{i,l}^+) \tag{17}$$

In the case where the variable is $x_i = y_i + c_i = 2^l(q_{i,l}^+ - q_{i,l}^-) + c_i$, formula $\|A\vec{x} - \vec{b}\|_2^2$ is used to express it as $\|A\vec{y} - (\vec{b} - A\vec{C})\|_2^2$ to change the global minimum energy being pursued [37]. Here, $y_i$ is a variable expressed as the superposition state of the qubit, and $c_i$ is a constant related to movement in search of the global minimum energy $-(\vec{b} - A\vec{C})^T(\vec{b} - A\vec{C})$.

The QUBO model, consisting of (12), (13), and (14), is a quantum-superposition state that can have any value in binary form for each $x_i$. To calculate an $n \times n$ linear system at the 64 bit level, more than $64n$ logical qubits are required. We introduce a two-dimensional example using (15), (16), and (17) to explain the algorithm that has an accuracy of 64 bits or higher with $3n$ qubits. In Fig. 1, circles and ellipses represent contour levels for $\|A\vec{x} - \vec{b}\|_2^2$. First, curves with eccentricity close to 0 at the contour level are considered, as in Figs. 1a, b, and c. The first step is to express $x_i$ in (10) using only the largest order. Then, $\vec{x}$ can be expressed as

$$\vec{x} \approx 2^m(q_{1,m}^+ - q_{1,m}^-, q_{2,m}^+ - q_{2,m}^-, \cdots, q_{n,m}^+ - q_{n,m}^-) \tag{18}$$

In Fig. 1a, the superposition state $\vec{x}$ can represent the corners of a square, center of an edge, or center of a face. The first step in the proposed algorithm is to calculate the QUBO model by substituting $l = m$ into (15), (16), and (17) corresponding to $\vec{x}$. In Fig. 1a, the origin represents the minimum energy among the nine points. The superposition state $\vec{x}$ of the second step is created around the minimum energy 0 obtained in the first step.

$$\vec{x} \approx 0 + 2^{m-1}(q_{1,m}^+ - q_{1,m}^-, q_{2,m}^+ - q_{2,m}^-, \cdots, q_{n,m}^+ - q_{n,m}^-) \tag{19}$$

The minimum energy is calculated by substituting $l = m - 1$ into (15), (16), and (17) to create the next QUBO model. The point corresponding to this minimum energy becomes the new center point in the next step. In Fig. 1b, $(2^{m-1}, -2^{m-1})$ becomes the minimum for the given energy and is the center point for the next steps. This procedure is repeated for the case of $l = m - 2$ using the new center point to find the solution (Fig. 1c). The proposed algorithm performs this task until it obtains a solution or converges.

The new algorithm may encounter problems when the eccentricity of the contour curve increases. Let us assume that the situation shown in Fig. 1d occurs at a certain step $l$.

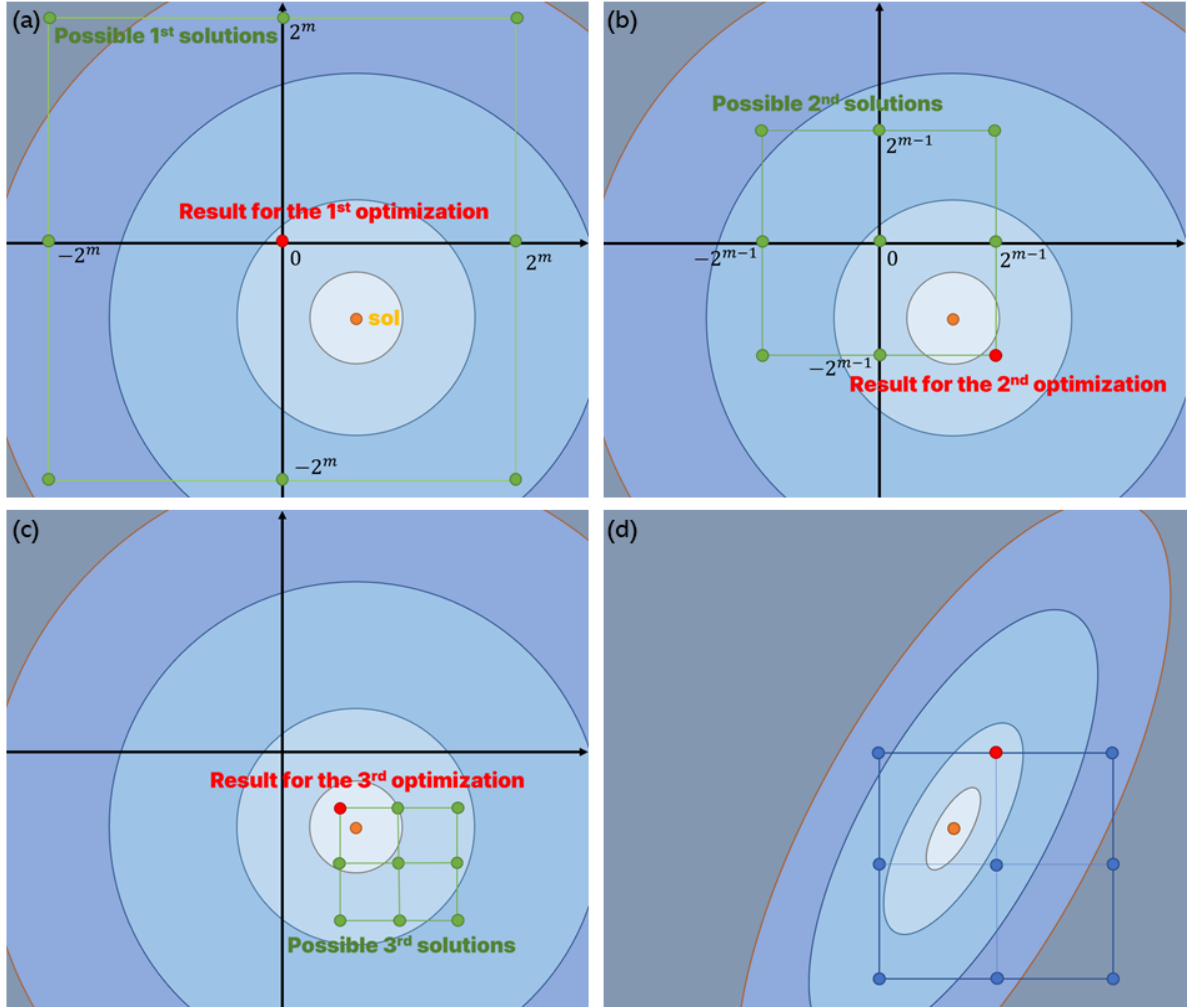

Figure 1. Iterative QUBO formulation algorithm. Finding the (a) minimum energy of the first step when the eccentricity of each curve of the contour for $\| A\vec{x} - \vec{b} \|_2^2$ is close to 0; (b) minimum energy of the second step centered on the minimum energy found in (a); (c) minimum energy of the third step centered on the minimum energy in (b). (d) The algorithm can obtain zero when the eccentricity $e$ of the curve is $0 < e < 1$ at a certain step.

Our algorithm moves $l-1$ steps around the point corresponding to the minimum energy in step $l$. A cube containing an overlapping $\vec{x}$ of level $l-1$ directs to a point far from the solution. This means that as we proceed through these steps, the algorithm may fail to reach a solution, or the solution found may contain algorithmic errors. To prevent this problem, the new algorithm adds an iterative operation that moves the center point at each step, creating a QUBO model in $l-1$ steps based on the optimization points found in step $l-1$ and finding the minimum energy for the new $3^n$ points in $n$-dimensional space. If the point directing to the new minimum energy is the same as the previously determined point, the algorithm proceeds to the next step, in which the size of the cube is reduced. The algorithm repeats the calculations until no new points are found. The overall algorithm is outlined in Fig. 2.

III. RESULT AND IMPLEMENTATION

In implementing the new algorithm in Python, we numbered the qubits in the order in which they are used to calculate the effective QUBO matrix. Because four qubits are used when calculating a $2 \times 2$ matrix, a QPU solver that can quickly perform numerous calculations was used [41]. As the QPU solver can accurately calculate up to 16 digits [30], this experiment used QUBO matrix coefficients of up to 17 digits.

The proposed algorithm was evaluated using the following linear system of irrational numbers (see Supplementary File 1):

$$\begin{bmatrix} \sqrt{2} & -\sqrt{3} \\ \sqrt{5} & \sqrt{7} \end{bmatrix} \vec{x} = \begin{bmatrix} 1024\sqrt{2}\pi + 32\sqrt{3}e \\ 1024\sqrt{5}\pi - 32\sqrt{7}e \end{bmatrix} \quad (20)$$

A total of 101 iterations were performed to obtain the solution. The first iteration was performed for $m = 20$, and $m$ was sequentially decreased by 1. Table 1 lists the number at which the QPU solver has the minimum value, and the difference between the positions of the solution and minimum energy has not changed for each of the three iterations in the QUBO model. The new algorithm increases precision by gradually reducing the minimum energy that the QUBO model must find and the values required for each bias and coupler. As shown in Table 1, with an increase in the number of iterations, the solution sought by the QUBO model gradually approaches the desired solution. For each $m$, the $(q_1, q_2, q_3, q_4)$ of the last experiment were (0, 0, 0, 0), so Table 1 shows the first experiment. In our experiment, the coefficient of $x_i$ was calculated from $2^{20}$ to $2^{-40}$, and the minimum energy that each QUBO model must find gradually decreased (See Fig. 3).

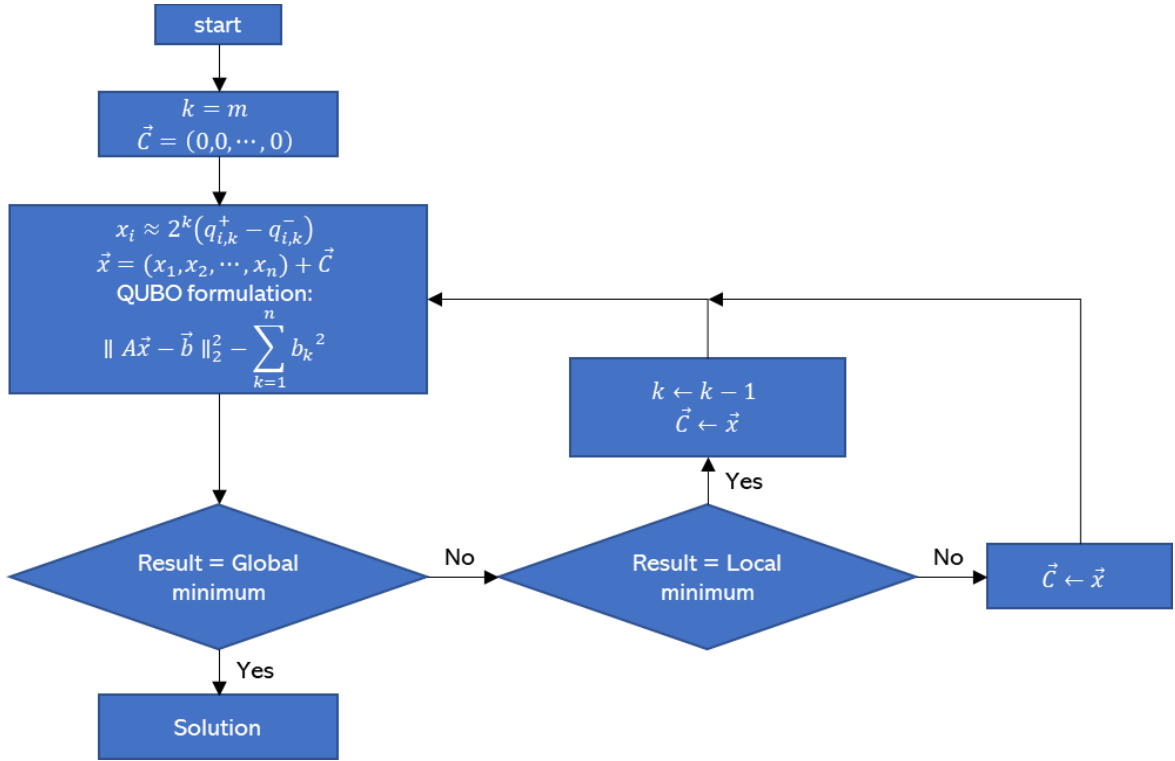

Figure 2. Workflow diagram for iterative QUBO formulation.

The second experiment was conducted on the same linear system, where each $x_i$ is $2^l(q_{i,l}^+ + 2q_{i,l+1}^+ + 4q_{i,l+2}^+ - q_{i,l}^- - 2q_{i,l+1}^- - 4q_{i,l+2}^-)$ using six qubits. Once the minimum energy that each QUBO model can express was determined, $l$ was decreased by 3. Through 39 iterations, we obtained the same solution as in the first experiment (see Supplementary File 2).

TABLE I. EXPERIMENTAL RESULTS FOR THE NEW ALGORITHM. THE NUMBER OF QUBIT COMBINATIONS OBTAINED WHEN THE STEP OF $\vec{x}$ IS EXPRESSED IN INTERVALS OF 5, THE NUMBER OF CASES WHERE THE MINIMUM ENERGY OCCURS, AND THE ERROR BETWEEN THE SOLUTION AT EACH STEP AND THE ACTUAL SOLUTION.

| $m$ | $(q_1, q_2, q_3, q_4)$ | #Occurrences | Error |
|---|---|---|---|
| 15 | (0,0,0,0) | 1000 | $3.22 \times 10^3$ |
| 10 | (0,1,0,0) | 1000 | $1.69 \times 10^2$ |
| 5 | (1,0,0,1) | 991 | $1.75 \times 10$ |
| 0 | (1,0,0,1) | 1000 | $1.75 \times 10^{-2}$ |
| -5 | (0,0,0,0) | 887 | $1.75 \times 10^{-2}$ |
| -10 | (1,0,0,1) | 980 | $4.71 \times 10^{-4}$ |
| -15 | (0,1,1,0) | 1000 | $3.33 \times 10^{-6}$ |
| -20 | (0,0,0,1) | 987 | $2.52 \times 10^{-7}$ |
| -25 | (0,0,0,1) | 936 | $1.21 \times 10^{-8}$ |
| -30 | (0,1,0,0) | 577 | $6.73 \times 10^{-10}$ |
| -35 | (0,0,0,1) | 752 | $1.11 \times 10^{-11}$ |
| -40 | (0,0,0,0) | 997 | $3.27 \times 10^{-13}$ |

## IV. DISCUSSION

Equations (10) and (11) indicate that when representing each $x_i$ in a superposition state, the number of required logical qubits increases proportionally with increasing accuracy. In the C and C++ programming languages, the double type can represent values up to $1.7E+/-308$. To simultaneously represent this in a quantum state, more than 1000 qubits are required. To create a quantum optimization model for an $n \times n$ linear system within a range similar to that of conventional methods, more than $1000n$ logical qubits are necessary. Furthermore, as the number of qubits in the binary representation increases, the difference between the maximum and minimum elements in the QUBO matrix also increases, making it challenging to determine the global minimum and maximum energies for the QUBO model. The proposed algorithm utilizes $3n$ logical qubits to compute linear systems, thereby providing a more accurate representation than classical methods. Furthermore, by using a hybrid solver that offers 1,000,000 variables, the new algorithm enables precise calculations for linear systems of up to $333,333 \times 333,333$.

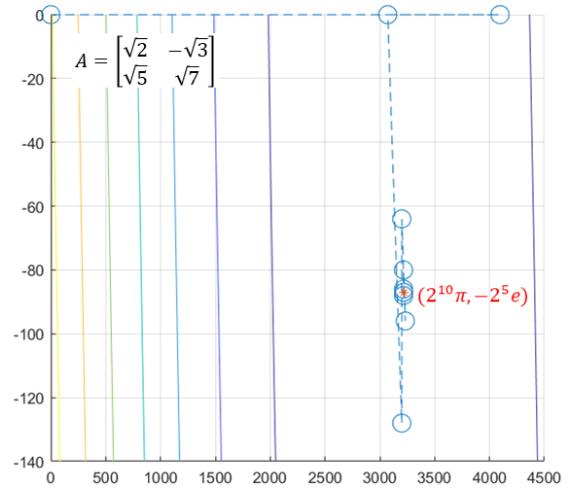

Figure 3. Movement of the center of the new algorithm. Each circle represents a center, the dotted lines show the movement of the center, and the red asterisk indicates the solution to the linear system.

The proposed algorithm was developed using only two qubits for each $x_i$. This algorithm would be even more effective if more qubits could be employed for each $x_i$. For example, if we use $x_i = 2^{m-2}(q_{i,m-2}^+ + 2q_{i,m-1}^+ + 4q_{i,m}^+ - q_{i,m-2}^- - 2q_{i,m-1}^- - 4q_{i,m}^-)$, the number of points that the superposition state $\vec{x}$ can represent becomes $15^n$. This can significantly reduce the difference between the points found at each step for the minimum energy and actual solutions when the eccentricity of the contour curve increases. We believe that an optimization process is required to determine the number of qubits used for each $x_i$, considering various factors such as the size of the problem and the time taken by the hybrid solver to compute the solution. In the case of using the QPU solver [36], reducing the number of qubits is the most crucial factor. However, the hybrid solver can quickly find solutions even with thousands of qubits [31]. This algorithm can adjust the total number of qubits used at each step, allowing consideration of the total qubits in terms of energy. We are currently conducting further experimental studies that consider these factors.

The new algorithm produces better results when the eccentricity of each curve of the contour for $\|A\vec{x} - \vec{b}\|_2^2$

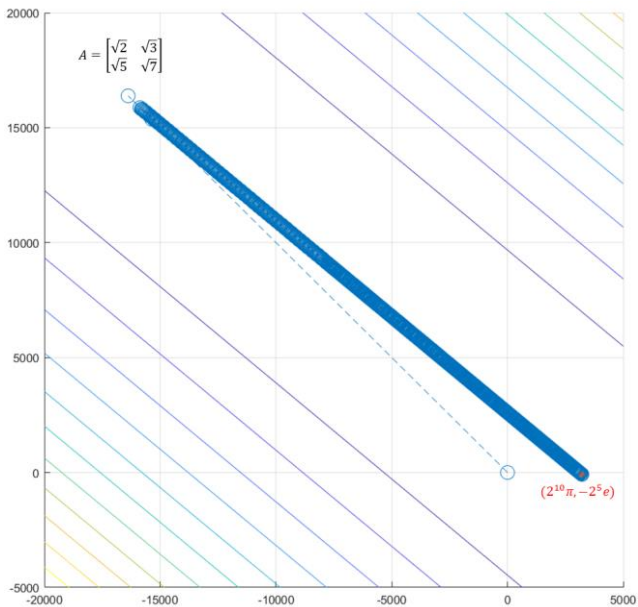

Figure 4. Example of inefficient iterative QUBO formulation algorithm. The red star is the solution of the linear system, and the two equilibrium lines of the same color represent each contour of $\|A\vec{x} - \vec{b}\|_2^2$. Each circle represents the center that the new algorithm is searching for at each step. *The dashed line indicates the movement of the center. This example may find points farther from the solution than in the previous step because the point of minimum energy at a particular step is the point with the minimum vertical distance on a straight line through the solution.*

approaches 0. In general, when the condition number of matrix $A$ increases, eccentricity also increases. Fig. 4 shows the application of the new algorithm when the condition number of the matrix is 129.44. The coefficient of $x_i$ from $2^{20}$ to $2^{17}$ points to the origin. Afterwards, when the coefficient is $2^{16}$, the minimum energy of the QUBO model is found to be $(-16384, 16384)$. This is because the distance from the straight line that is parallel on each contour and includes the solution is closer to point $(-16384, 16384)$ than the origin. In the new algorithm, the coefficient of $x_i$ decreases rapidly at each step, requiring many iterations to find the solution. The new algorithm requires more calculations as the value of the eccentricity approaches 1. The proposed algorithm can always determine a solution if there exists one when representing $\vec{x}$ as a superposition. However, if the superimposed $\vec{x}$ does not contain a solution, the new algorithm calculates a point using global minimum energy.

Knowing the eigenvalues and eigenvectors, allows $\vec{x}$ to be used more efficiently than in (10) and (11). Let the eigenvalues and the unit eigenvectors be $\{\lambda_1, \lambda_2, \cdots, \lambda_n\}$ and $\{\vec{v_1}, \vec{v_2}, \cdots, \vec{v_n}\}$, respectively. Subsequently, $\vec{x}$ can be expressed as follows:

$$\vec{x} = \sum_{i=1}^{n} \left( \sum_{l=-m}^{m} 2^l q_{i,l}^+ - \sum_{l=-m}^{m} 2^l q_{i,l}^- \right) \vec{v_i} \qquad (21)$$

The movement at each step is in a direction perpendicular and horizontal to the contour, which is expected to be more efficient when eccentricity is close to 1 because it prevents the center from moving in a zigzag manner, as shown in Fig. 3. We expect that with the development of quantum computers/annealers, this new algorithm will be widely used in many fields.

DATA AVAILABILITY

This algorithm was developed on July 26, 2022, and the source code can be found at the GitHub address below:

https://github.com/ktfriends/Numerical_Quantum_Computing/tree/main/HP

ACKNOWLEDGMENT

This research used resources of 'Quantum Information Science R&D Ecosystem Creation' through the National Research Foundation of Korea(NRF) funded by the Korean government (Ministry of Science and ICT(MSIT))(No. 2020M3H3A1110365). H. L. is supported by the National Research Foundation of Korea(NRF) grant funded by the Korea government(MSIT)(RS-2024-00352408). K. J. is supported by the MSIT(Ministry of Science and ICT), Korea, under the ITRC(Information Technology Research Center) support program(IITP-RS-2024-00437284) supervised by the IITP(Institute for Information & Communications Technology Planning & Evaluation).